\documentclass[aps,manuscript]{revtex4-2} 
\usepackage{bm}
\usepackage{amsmath}
\usepackage{color}
\usepackage{graphicx}
\usepackage{amssymb}
\usepackage[toc,page]{appendix}

\DeclareMathAlphabet{\mathbbmsl}{U}{bbm}{m}{sl}
\makeatletter
\newsavebox{\@brx}
\newcommand{\llangle}[1][]{\savebox{\@brx}{\(\m@th{#1\langle}\)}%
	\mathopen{\copy\@brx\kern-0.5\wd\@brx\usebox{\@brx}}}
\newcommand{\rrangle}[1][]{\savebox{\@brx}{\(\m@th{#1\rangle}\)}%
	\mathclose{\copy\@brx\kern-0.5\wd\@brx\usebox{\@brx}}}
\makeatother
\begin{document}
\draft

\title{Magnetoplasmons in magic-angle twisted bilayer graphene}

\author{Thi-Nga Do$^{1}$, Po-Hsin Shih$^{2}\footnote{Corresponding author: {\em E-mail}: shih.pohsin@gmail.com}$, Godfrey Gumbs$^{2,3}\footnote{Corresponding author: {\em E-mail}: ggumbs@hunter.cuny.edu}$}
\affiliation{$^{1}$ Department of Physics, National Cheng Kung University, Tainan 701, Taiwan  \\
$^{2}$ Department of Physics and Astronomy, Hunter College of the City University of New York,
695 Park Avenue, New York, New York 10065, USA \\
$^{3}$ Donostia International Physics Center (DIPC), P de Manuel Lardizabal, 4, 20018 San Sebastian, Basque Country, Spain \\}

\date{\today}

\begin{abstract}

The magic-angle twisted bilayer graphene (MATBLG) has been demonstrated to exhibit exotic physical properties due to the special flat bands. However, exploiting the engineering of such properties by external fields is still in it infancy.
Here we show that MATBLG under an external magnetic field presents a distinctive magnetoplasmon dispersion, which can be significantly modified by transferred momentum and charge doping.
Along a wide range of transferred momentum, there exist special pronounced single magnetoplasmon and horizontal single-particle excitation modes near charge neutrality. We provide an insightful discussion of such unique features based on the electronic excitation of Landau levels quantized from the flat bands and Landau damping.
Additionally, charge doping leads to peculiar multiple strong-weight magnetoplasmons. These characteristics make MATBLG a favorable candidate for plasmonic devices and technology applications.

\end{abstract}
\maketitle

\section{Introduction}
\label{sec1}

Carbon-based two-dimensional (2D)  materials have attracted considerable attention from both theoretical and experimental scientists due to their unique physical properties. For twisted bilayer graphene (TBLG) at the magic angles, the two flat bands near charge neutrality lead to strong electronic coupling, giving rise to intriguing physical phenomena such as correlated insulator \,\cite{insulator1, insulator2, insulator3}, superconductor \,\cite{insulator2, insulator3}, ferromagnetism \,\cite{ferromagnetism}, charge-ordered states \,\cite{charge}, as well as quantized anomalous Hall effect \,\cite{QAHE}. The band structures of TBLG at large angles and magic angles have been experimentally probed \,\cite{bandexp, bandexp2} and predicated by density-functional theory \,\cite{DFT1, DFT2} and modeling\,\cite{Mac, Koshino, effective, 8band}.   So far, the magnetic quantization of TBLG has been extensively studied to identify the Landau levels (LLs) \,\cite{LLexp1, LLexp2} and quantum Hall conductivity (QHC) \,\cite{QHC1, QHC2, filling}. For magic-angle TBLG (MATBLG), LL degeneracy variation associated with filling of electrons has been observed by transport measurements \cite{insulator1, insulator2, insulator3, filling}. Apparently, TBLG is a leading candidate for potential applications in new-generation devices with modern advanced functionalities\,\cite{app}.

\medskip
\par
Electron correlations in flat-band systems such as MATBLG offer insight into fundamental physical phenomena of the materials. The connection between the electronic correlations and quantum degeneracy enables a thorough explanation for the microscopic mechanism of the unique electronic phases in MATBLG, including topological phases \cite{8band4, topo1}, correlated insulators \cite{insulator1, insulator2, insulator3}, and superconductors \cite{insulator2, insulator3}. Spectroscopy observation of strong electron correlations in partially-filled flat bands has been reported previously \cite{insulator1, charge}. Up to now, collective excitation of MATBLG has been investigated with the use of optical microscopies \cite{plasmonexp} and predicted by theoretical calculations \cite{plasmontheo}. The plasmon modes exhibit intriguing characteristics, including relatively high group velocity and enhanced role of electron correlations.

\medskip
\par
Collective charge density fluctuations have been demonstrated to be significantly enriched by the application of an external magnetic (B) field. The excitation between quantized LLs gives rise to the so-called "magnetoplasmons" which are self-sustained charge density oscillations. The concept of magnetoplasmon is defined as the collective excitation between B-field-quantized LLs. Such phenomenon occurs for sufficiently strong B fields, which is distinct from the zero-field case. So far, several experimental and theoretical studies on the plasmon of TBLG in the absent of a B field have been reported in the literature \,\cite{plasmonexp, R14, R15}. Infrared-optical absorption and inelastic-light scattering\,\cite{infrare1, infrare2} are well-known techniques for investigating magnetoplasmons in condensed matter systems. Specifically, magnetoplasmon for graphene has been successfully measured experimentally \,\cite{expgraphene}. On the theoretical side, magnetoplasmons have been investigated in layered graphene \,\cite{layered,GG1,GG2}, doped graphene \,\cite{doped}, graphene nanoribbons \,\cite{GNR}, and silicene\,\cite{sifield}. Magnetoplasmonics play an important role in technological applications \,\cite{tech1, tech2}.

\medskip
\par
It is expected that the influence of an external B field on the collective excitations of MATBLG leads to novel physical phenomena, which are worthy of thoroughly investigating. In this paper, such physical phenomena will be addressed by employing the tight-binding model (TBM) combined with the Peierls substitution and the dielectric function.
The simplified N-band models (N is typically in the range 2-10) have been reported previously \cite{8band3, R17, R18}. These models do not take into account the spin and momentum-space valley degrees of freedom. They are chosen to address the flat bands which arise near the magic angle while taking advantage of the roughly 30-meV band gaps between the flat bands and other bands. This assumes that the correlations responsible for the observed superconductivity can be accurately described with just the flat bands. If the interaction strength in MATBLG is larger than the gaps that separate the flat band manifold, a model which includes additional nearby bands is required. Since the LLs are quantized from the electronic bands, all calculations related to LLs must be based on a reliable energy band structure. Therefore, it is crucial to build up a suitable model which can explicitly preserve all symmetries of the full TBM like the 8-band model which we present in this work \cite{8band}.
We will show the exotic magnetoplasmons in MATBLG and their significant dependence on electron doping and transferred momentum.

\section{Theoretical Method}
\label{sec2}

\subsection{Minimal-basis tight-binding model and Peierls substitution}

When dealing with problems in theoretical physics, one comes to realize that determining the eigenstates of the Hamiltonian is the pre-eminent task for any further investigation of the fundamental physical properties of the material. Hamiltonian matrices of many condensed matter systems can be solved efficiently by computational methods. Limitation of the numerical technique is a great challenge in dealing with a large size Hamiltonian of a system such as MATBLG.  In this work, we employ the minimal-basis 8-band TBM, which enables the study of electron correlation phenomena in real situations \cite{8band}. The tight-binding Hamiltonian for MATBLG system can be written as

\begin{equation}
H  = \sum\limits_{m,m'} t_{mm'}(c_{m})^\dagger c_{m' } + H.c..
\end{equation}
In this notation, $m$ and $m'$ specify the effective lattice sites, $t_{mm'}$ stands for the hopping terms which describe the atomic interactions. The annihilation operator $c_{m}$ (creation operator $(c_{m})^\dagger$) can destroy (generate) an electronic state at the effective $m$-th site. The . H.c. denotes the Hermitian conjugate.

\medskip
\par
\begin{figure}[htbp]
\begin{center}
\includegraphics[width=0.9\linewidth]{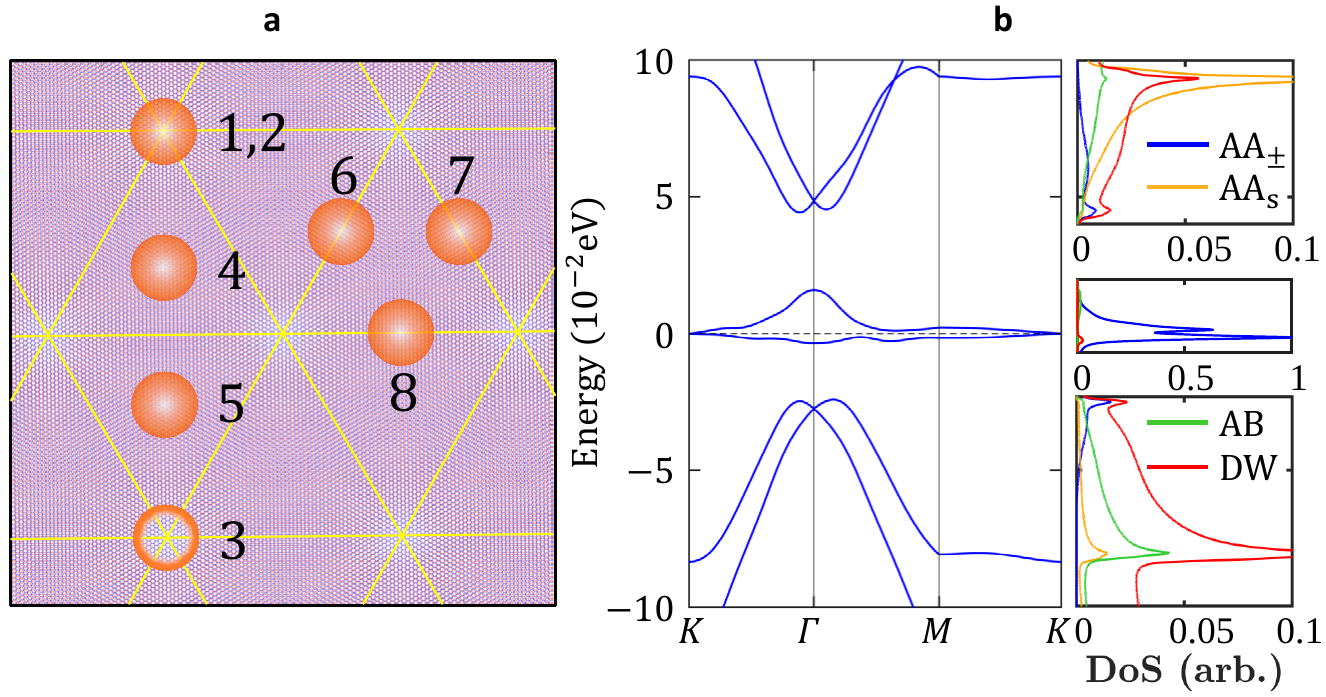}
\end{center}
\caption{(color online) Lattice structure, band structure, and DOS of TBLG at $\theta = 1.1^{\circ}$. The circles in {\bf a} show the eight trial Wannier functions following the symmetry prescription. Plot {\bf b} captures six low-lying bands and corresponding DOS with energy-dependent contribution of four atomic orbitals $AA_{\pm}$, $AA_{s}$, $AB$, and $DW$. }
\label{Fig1}
\end{figure}

The minimal-basis TBM is generated based on the combination of a proper {\em ab initio\/}  $\mathbf{k.p}$ continuum model Hamiltonian and a multi-step Wannier projection technique. Here, we employ an eight-band model from Ref. \cite{8band}, which consists of two flat bands near the Fermi level, $E_F$, and three upper bands together with three lower bands. This model describes TBLG by 8 high-symmetry points corresponding to various orbitals, as depicted in Fig.\,\ref{Fig1}a. The contributions due to each orbital to the eight bands near $E_F$ are unequivalent. There is a great number of parameters, which are generated by using a multi- step Wannier projection technique. These parameters are available as part of a larger collection of continuum models for twisted bilayer graphene by Carr et al. in Ref. \cite{R1}. The 8-band model Hamiltonian is an 8$\times$8 matrix with the basis built up by 8 trial Wannier functions. Each matrix element is a diagonal block with size depending on the number of hopping parameters. Here, we consider the orbital interactions up to 10-nearest neighboring sites along 4 directions, $\pm \hat{x}$ and $\pm \hat{y}$. As a result, each block is as large as 20$\times$20. Therefore, the 8-band model Hamiltonian is actually a 160$\times$160 matrix with 25,600 hopping parameters.

\medskip
\par
For MATBLG, there exist the following candidate effective lattice sites: (1) triangular, labeled $\tau$ and corresponding to AA stacking regions, (2) honeycomb, labeled $\eta$ and corresponding to AB/BA stacking regions, and (3) kagome, labeled $\kappa$ and corresponding to an intermediate stacking regions between AB and BA sites. For TBLG with small angle $\theta <$ 1.5$^{o}$, the honeycomb sites become large triangular domains of AB or BA stacking, and the kagome lattice locations can be interpreted as domain walls (DW) between them. The four possible orbital symmetries, s, $p_z$, and $p_{\pm}$ ($p_{\pm} = p_x \pm ip_y$) correspond to the same symmetry eigenvalues expected of a hydrogen orbital with the same label. The symmetries and interpretations of the 8 Wannier functions presented in Fig. 1a are shown in Table 1.

\medskip
\par
The effects of atomic reconstruction due to relaxations play an important role in the electronic properties of materials. The relaxations in MATBLG, including the in-plane and out-of-plane ones, are taken into account in our TBM \cite{8band}. In-plane relaxation decreases the area of the high stacking energy AA region while it increases that of low stacking energy AB/BA regions. Out-of-plane relaxation causes corrugation, increasing the vertical separation between the AA regions from the equilibrium distance in AB stacking. It is worth noting that the electron density that forms on the AA stacking sites, which is associated with the $AA_{\pm}$ orbitals, near the charge neutrality point is responsible for the most important interaction terms for TBLG's flat bands. That is, atomic relaxations result in the band gaps either side of the flat bands. This effect has also been discussed in previous studies by various groups \cite{DFT1, 8band, relaxation1, relaxation2}.

\medskip
\par
The 8-band model is a trustworthy model since necessary response tests have been carried out and it turned out that all symmetries of the full TBM are explicitly preserved. The wave functions have underlying index polarization and complex phase consistent with their symmetry descriptions. Furthermore, it captures most of the features in the $\mathbf{k.p}$ model such as the effect of the layer-exchanging symmetries breaking, the flat-band manifold, and the relevant band gaps. In addition, the major parts of model and resultant band structure are confirmed by other experimental and theoretical studies \cite{8band1, 8band2, 8band3, 8band4, 8band5}. The 8-band model has an advantage over existing methods because on one hand, it preserves all symmetries which are included in the full TBM. However, on the other hand, it connects to realistic {\em ab  initio\/} band structures without any band fitting procedures. Furthermore, the inclusion of other bands in the vicinity of the charge neutrality point in addition to the flat bands allows for the full capture of critical physical features of the MATBLG.

\medskip
\par
\begin{table}[h]
  \begin{center}
    \caption{Orbitals of the reduced Hamiltonians. The Wannier function (WF) index, lattice description, and symmetry properties
for the eight-band model are given in the first three columns. The symbols $\tau$, $\eta$, and $\kappa$ correspond to a
triangular, hexagonal, and kagome lattices, respectively. The symbols
$p_{\pm}$, $p_z$, and s refer to the symmetries of a hydrogenic wave function. The fourth column gives a description
of the WF character, including relations to stacking order and any layer or sublattice polarization. The last column shows the notation of orbitals presented in Fig. 1a.}
    \label{tab:table1}
    \begin{tabular}{|c|c|c|c|c|}
      \hline
      \textbf{WFs} & \textbf{Lattice} & \textbf{Symmetry} & \textbf{Orbital description} & \textbf{Orbital notation}\\
      \hline
      1 & $\tau$ & $p_{+}$& B orbital dimers at AA& $AA_+$ \\
      2 & $\tau$ & $p_{-}$& A orbital dimers at AA& $AA_-$ \\
      3 & $\tau$ & s& Equal mixture around AA& $AA_s$ \\
      4 & $\eta$ & $p_{z}$& B on $L_1$, A on $L_3$ at BA& AB \\
      5 & $\eta$ & $p_{z}$& A on $L_1$, B on $L_2$ at AB& AB \\
      6 & $\kappa$ & s& BA(AB) left (right) of DW& DW \\
      7 & $\kappa$ & s& AB(BA) left (right) of DW& DW \\
      8 & $\kappa$ & s& AB(BA) above (below) DW& DW \\
      \hline
    \end{tabular}
  \end{center}
\end{table}

\medskip
\par
To consider the effect due to a uniform external B field, we developed a new method which combines the minimal-basis TBM and the Peierls substitution.
Under a uniform perpendicular B field $\mbox{\boldmath$B$}= (0,0,B)$, the lattice periodicity is noticeably modified. Thus, we include the Peierls phase $G_{R} = (2\pi/\phi_{0})\int\limits_{\bf R}^{\bf r} \mbox{\boldmath$A$}\cdot d\mbox{\boldmath$\ell$}$ in the Hamiltonian by using the Peierls substitution\,\cite{Pei1933, PRBShih}. Here, ${\bf R}$ and ${\bf r}$ refer to the position vectors of the unit cells within the Bloch basis which are defined as the lattice sites and layers $m,m';\,\ell,\ell^{\prime}$. $\mbox{\boldmath$A$}=(0,Bx,0)$ is the magnetic vector potential in the Landau gauge and $\phi_{0}=hc/e$ is the flux quantum. The B field induces the unit cell expansion along the x direction which makes an angle of $\theta/2$ with the armchair direction of the untwisted graphene.
The number of atoms in an extended cell, $2N \times 2\phi_{0}/ \phi$, is defined by that of a primitive unit cell ($N$) and a period of the Peierls phase ($2\phi_{0}/ \phi$). In this notation, $\phi = B{\cal S}$ is the magnetic flux and ${\cal S}$ is the area of a unit cell.
Usually, the ratio of flux through a lattice cell to one flux quantum, $\alpha = \phi/\phi_0$, is assumed to be irrational so that the Bloch band always breaks up into precisely $\phi_0$ distinct energy bands (LLs). However, it was shown by Hofstadter \cite{R20} that the B-dependent LL energy spectrum establishes a totally continuous behavior for all magnetic field values.
It is worth noting that for the minimal-basis TBM, the Bloch basis of wave functions are employed. Within Bloch basis, only the distance between unit cells matters since each unit cell is viewed as a point \cite{PRBShih}. Therefore, minimal-basis TBM can avoid the phase-shift in y direction when the magnetic field is taken into consideration.
The magnetic Hamiltonian becomes

\begin{equation}
H_B =\sum\limits_{m,m'}\, t_{mm'}\,e^{iG_R}\,\left(c_{m}\right)^\dagger c_{m'} + H.c..
\end{equation}

Although there is a huge number of Peierls phases for MATBLG, the Peierls substitution method is still applied properly and thus the results achieve a high level of accuracy as for monolayer graphene \cite{PRBShih}. The main features of quantized LLs can be well explained and built from the B = 0 band structure. The eigenvalues and wave functions obtained by solving the magnetic Hamiltonian matrix are then inserted into the dielectric function in order to calculate the electron excitation phase diagrams.

\subsection{Dielectric function}

Under an impinging electron beam, the electrons of TBLG can be subjected to charge redistribution due to the frictional force created by the external electric field. In response to this external electric field associated with this beam, the electrostatic potential of the interacting electrons is screened through a dielectric function which is determined by a polarization propagator.
The theoretical method which combines dielectric function, either static or dynamic form, with material models like TBM and continuum model has been commonly used by different groups to investigate the electric excitation of TBLG and other materials.
The calculated static random-phase approximation (RPA) and constrained RPA polarizability of undoped TBLG allowed for the understanding of dielectric response of the materials and its dependence on various factors such as internal screening, dielectric environment, and twist angle \cite{dielectric1, dielectric2}.
In addition, the dynamical response of a material to an outside electric perturbation was explored by employing the dynamic dielectric function. Some interesting phenomena were found for TBLG at a small twisted angle or magic angle, including the collective excitonic in-phase oscillations \cite{dielectric3}, intrinsically undamped plasmon modes \cite{dielectric4}, and dependence of energy loss function on the filling factors and twist angles \cite{R14}.

\medskip
\par
For MATBLG, the interlayer coupling is significant, thus the contribution of each layer to the band structure can not be classified. Furthermore, the dielectric function can only be defined for the whole system but not for individual layers. Thus, we treat this problem of magic-angle TBLG as for a single-layer system. The dielectric function, within the RPA is given by \cite{R4}

\begin{equation}
\epsilon(\mathbf{q},\omega) = \epsilon_1(\mathbf{q},\omega) + i\epsilon_2(\mathbf{q},\omega) = \epsilon_0 - V_q\chi^0(\mathbf{q},\omega).
\end{equation}
In this notation, $\epsilon_0 = 2.4$ is the background dielectric constant of graphite. The choice of $\epsilon_0$ does not affect the quantitative picture of the plasmon dispersion. Also, $\omega$ is the angular frequency of a testing field, $\mathbf{q}$ is the momentum transfer (with magnitude $q$), and $V_q = 2\pi \frac{e^2}{q}$ presents the in-plane Fourier transformation of the bare Coulomb potential energy. $\chi^0(\mathbf{q},\omega)$ is the 2D bare response function which can be written as \cite{R4}

\begin{equation}
\chi^0(\mathbf{q},\omega)=\frac{1}{S}\sum_{m,m'} | \langle m, \mathbf{k} +  \mathbf{q} |e^{i\mathbf{q}.\mathbf{r}}| m', \mathbf{k} \rangle|^2 \\
\times \frac{f(E_m) - f(E_{m'})}{E_m - E_{m'} - (\omega - i\Gamma)}.
\end{equation}
Here, $S = 3\sqrt{3}b^2\frac{2\phi_0}{\phi}$ is the area of the B-field-extended unit cell. The normalization constant $1/S$ ensures the equal contribution of each $\mathbf{k}$ state in the first BZ. $\Gamma = 0.1$ meV is the dephasing factor due to deexcitation mechanisms. $E_{m,m'}$ are the energies of the LLs involved in the excitation. $f(E) = \frac{1}{1 +\exp[(E - \mu_T)/k_BT]}$ is the equilibrium Fermi-Dirac distribution function, in which $k_B$ is the Boltzmann constant, T is temperature and $\mu_T$ is the T-dependent chemical potential. In this work, we consider the zero temperature situation. Note that for low T, the chemical potential $\mu_T$ is defined by \cite{R5}

\begin{equation}
\mu_T = \mu_0 \left\{ 1- \frac{\pi^2}{6} \frac{\textrm{d log D}(\mu_0)}{\textrm{d log}\mu_0} \left(\frac{kT}{\mu_0}\right)^2  + ... \right\}
\end{equation}
where $D(E)$ is density-of-states, $\mu_0$ is the Fermi energy. Since T is generally small compared with the Fermi temperature $\mu_0/k_B$,  the $T^2$ term is much smaller and thus its effect at low temperatures on $\mu_T$ is not significant. Up to now, there have been many studies on the plasmon dispersion at zero temperature \cite{GG2, R6, R7, R8, R10}. Interestingly, experimental measurements \cite{R11} have verified the square root plasmon dispersion at finite T for graphene which is in  excellent agreement with theoretical predictions \cite{R6}.

\medskip
\par
In Eq. (6), the term $\langle m, \mathbf{k} +  \mathbf{q} |e^{i\mathbf{q}.\mathbf{r}}| m', \mathbf{k} \rangle$ represents the Coulomb matrix elements which can be written as

\begin{equation}
\langle m, \mathbf{k} +  \mathbf{q} |e^{i\mathbf{q}.\mathbf{r}}| m', \mathbf{k} \rangle =
\sum_{n,n'}\sum_{m,m'}\langle \phi_z(\mathbf{r} - \mathbf{R}  ) |e^{-i\mathbf{q}.(\mathbf{r} - \mathbf{R})}|   \phi_z(\mathbf{r} - \mathbf{R}  )  \rangle [u_{n,m}(\mathbf{k} +  \mathbf{q}) u^{*}_{n',m'}(\mathbf{k}) ],
\end{equation}
where $n$ and $n'$ denote the LL indexes, and $\mathbf{k} $ and $\mathbf{k} +  \mathbf{q}$ are wave vectors. The tight-binding function is defined as a superposition of the product of the coefficient $u_{n,m}(\mathbf{k} +  \mathbf{q})$ (also known as the subenvelope function) and the position-dependent 2$p_z$ orbital function of carbon atoms $\phi_z$. Theoretically, the subenvelope functions can be obtained by diagonalizing the Hamiltonian matrix. However for the huge Hamiltonian matrix as for MATBLG, they are the outcome of high-technique simulation calculations. Furthermore, $\phi_z$ is approximated as a generalized hydrogenic wave function of the form $\phi_z(r) = Arcos(\theta)e^{-Zr/2a_0}$ with $A$ the normalization factor,  $\theta$ the angle from the $c$ axis, $a_0$ the Bohr radius, and $Z = 3.18$ an effective core charge. This yields $\langle \phi_z(\mathbf{r} - \mathbf{R}  ) |e^{-i\mathbf{q}.(\mathbf{r} - \mathbf{R})}|   \phi_z(\mathbf{r} - \mathbf{R}  )  \rangle = [1 + (qa_0/Z)^2]^{-3}$.

\subsection{Effect of screened potential due to doping}

To consider the effect of electron and hole doping on the electronic excitation of MATBLG, we include the doping-induced screened potential in the Hamiltonian. The tight-binding Hamiltonian becomes
\begin{equation}
H  = \sum\limits_{m,m'} t_{mm'}(c_{m})^\dagger c_{m' } + H_{doped} + H.c..
\end{equation}
Here, $H_{doped}$ stands for the doping-induced potential term which can be written as \cite{scpotential}
\begin{equation}
H_{doped}  = \sum\limits_{m} (U_{E_F,m} - U_{E_F=0,m}) (c_{m})^\dagger c_{m},
\end{equation}
where $U_m = Re[\sum\limits_{\mathbf{q}} V(\mathbf{q}) e^{i\mathbf{q}.\mathbf{R}_m}]$ is the potential energy for the $m$ orbital basis. In this notation, $V(\mathbf{q})$ is the Fourier-transformed potential, $V(\mathbf{r}) = 1/(4\pi\epsilon \mathbf{r})$, given by
\begin{equation}
V(\mathbf{q}) = \frac{e^2}{4\pi\epsilon(\mathbf{q})} \int d^2\mathbf{r} \frac{1}{\mathbf{r}} e^{i\mathbf{q}.\mathbf{r}} = \frac{e^2}{2\epsilon(\mathbf{q}) \mathbf{q}}
\end{equation}
with $\epsilon(\mathbf{q})$ being the statistic dielectric function in Eq. (3) at $\omega = 0$.

\section{Results and Discussion}
\label{sec3}

The energy band structure of TBLG at a magic angle $\theta = 1.1^{\circ}$ at low energies is presented in the left panel of Fig.\,\ref{Fig1}b. Our eight-band model yields eight energy bands. However, here we consider six low-energy bands which are associated with the low-frequency electron excitations. Each band is dominated by different sets of orbitals, which is clearly revealed in the density of states (DOS), referring to the right-most panel of Fig.\,\ref{Fig1}b. The two nearly flat bands near zero energy mainly come from the $AA_{\pm}$ orbitals. For MATBLG, the bandwidth of the flat bands is extremely narrow, giving rise to a large DOS as well as strong electron correlations.  It is well known that strong correlations occur if the interaction is large compared to the band width. In a flat-band system like MATBLG, the latter vanishes. Therefore, an arbitrary small interaction is already sufficient to produce strong correlation effects. Flat bands are advantageous because they guarantee a large DOS, which amplifies the effects of interactions. For the other four conduction and valence bands, the electronic states in the vicinity of the extremal points are dominated by the $AA_{\pm}$ and $DW$ orbitals, whereas the states elsewhere are governed by the combination of ($AA_{s}$, $AB$, $DW$). The DOS of the two nearly flat bands presents very high peaks, which are an order larger than those of the other bands. This is consistent with the characteristics of the flat bands.

\medskip
\par
\begin{figure}[htbp]
\begin{center}
\includegraphics[width=0.9\linewidth]{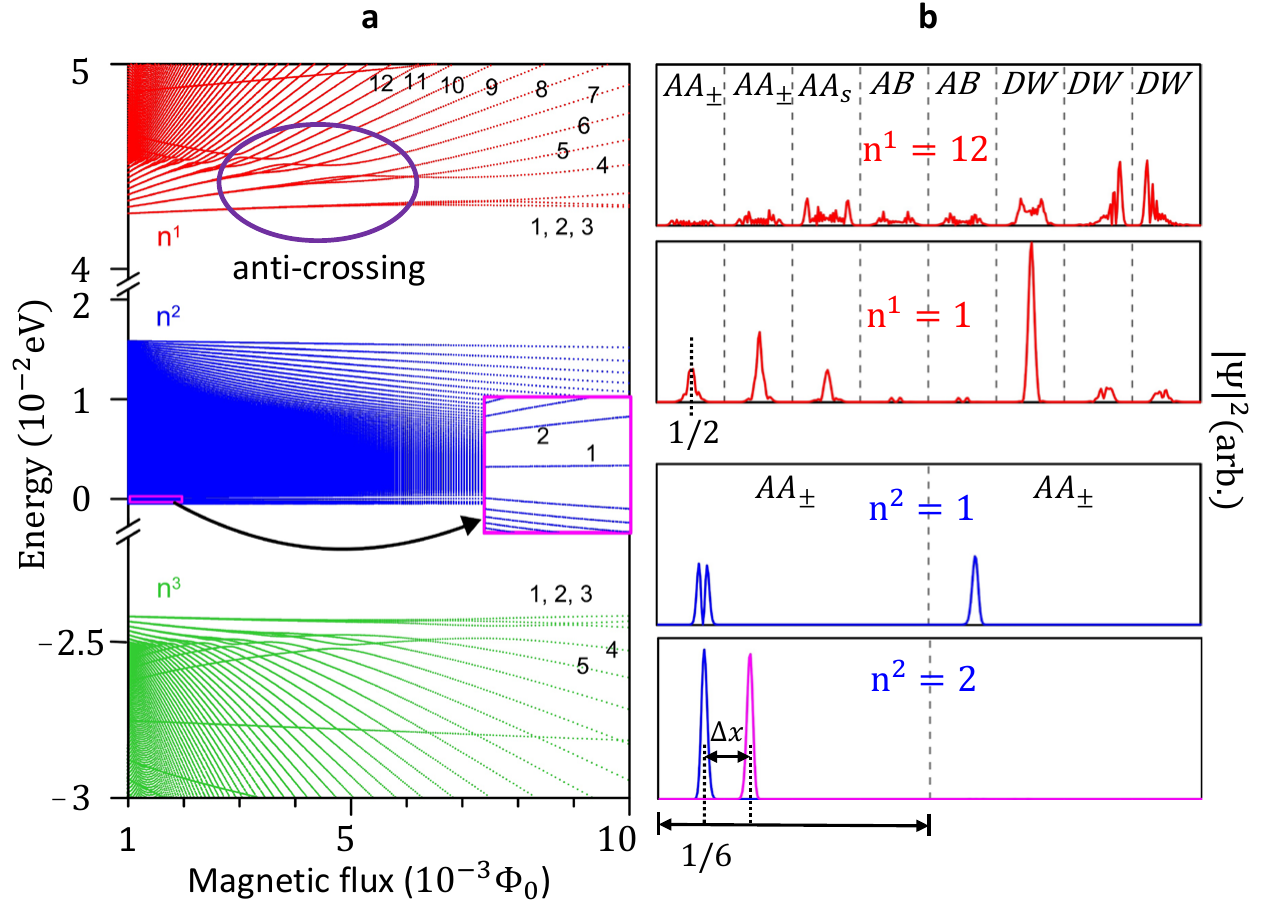}
\end{center}
\caption{(color online) LL energy spectrum and wave function amplitudes of TBLG at magic angle $\theta = 1.1^{\circ}$ for a flux $\Phi = 0.005\Phi_0$. {\bf a,} There are three groups of LLs and they are denoted by $n^1$ (red), $n^2$ (blue) and $n^3$ (green). The LL indexes are shown as integers for the first few LLs of each group. {\bf b,} The orbital distribution on each LL is non-equivalent. The $n^2$ LLs are mainly governed $A_{\pm}$, while all four orbitals are involved in $n^1$ LLs. Furthermore, the orbital distribution on $n^1 = 12$ becomes more equivalent than that on $n^1 = 1$. This can also be seen from the DOS in Fig. 1b. $\Delta x$ represents the shift of wave function nodes along the x direction, corresponding to the finite transferred momentum.}
\label{Fig2}
\end{figure}

The electronic states are quantized into three groups of LLs, as illustrated in Fig.\,\ref{Fig2}a. They are well separated by wide gaps, corresponding to the separation of three zero-field energy band groups. In the vicinity of charge neutrality, the $n^2$ LLs are quantized from the two nearly flat bands which have a finite number of electronic states within a narrow energy range. Consequently, there exists a finite number of $n^2$  LLs, which are located within the same energy range as the zero-field two flat bands. Furthermore, the very high density of $n^2$ LLs is consistent with the high DOS of the two flat bands. The B-dependent $n^2$ LL spectrum is similar to that of monolayer graphene with a dispersionless LL at E = 0 and a nearly linear-dispersion LL in the surrounding region (see the zoom-in of Fig.\,\ref{Fig2}a). On the other hand, the $n^1$ and $n^3$ groups exhibit the anti-crossing phenomenon. According to the Wigner-von Neuman non-crossing rule, two LLs simultaneously possessing at least an identical wave function mode with comparable amplitude on specific sublattices cannot cross each other. Instead, they are split apart at the so-called ``avoided crossing point". In general, the energy range of quantized LLs corresponds to that of the zero-field energy bands. Note that the other bands, which are not included in the minimal model, are located at higher and deeper energies. So do their quantized LLs. Thus, these LLs have no contribution to the electronic excitations in the energy range we consider. In fact, the LLs and QHC of the MATBLG based on the 8-band model have been carried out in our previous study \cite{PRBShih}. The calculated results have been shown to be consistent with theoretical and experimental reports, regarding the LL degeneracy and QHC plateaus.
It is worth mentioning that the magnetic quantization can lead to a Hofstadter butterfly if the B field satisfies specific conditions based on the interplay between lattice periodicity and magnetic length \cite{R20, R21, fragiletopo}. The Hofstadter butterfly of TBLG shows some interesting phenomena such as the nonmonotonic behavior of the Hall conductivity as a function of Fermi energy \cite{R21} and the connection between the fragile topological bands and other bands \cite{fragiletopo}.
In general, butterfly states emerge when the periodicity becomes comparable with the magnetic length. For  MATBLG, the Hofstadter butterfly starts to appear when the flux ratio satisfies $\phi/\phi_0 \geq 0.2$ \cite{R21}. Here, this is not the case in our work. We consider the comparatively  weak B field with $\phi/\phi_0 \leq 0.01$ in order to focus on the quantization of the flat bands. Therefore, there is no appearance of a Hofstadter butterfly for our considered range of the B-field.

\medskip
\par
The behavior of LLs can be understood based on the coefficients of the tight-binding functions, i.e., the sub-envelope functions. Figure\,\ref{Fig2}b illustrates the distribution probability for the first few LLs of the $n^1$ and $n^2$ groups. They feature the characteristics of the Hermite polynomial functions. As known, Hermite polynomials are a set of polynomial functions used to describe the wave functions for the Hamiltonian of a quantum harmonic oscillator. They possess a number of characteristics such as even/odd symmetry, orthogonality, and recurrence relation \cite{R12}. This is the key point for understanding the inter-LL excitations based on the Coulomb matrix elements of the dielectric function. The $n^2$ LL wave functions show the dominant well-behaved nodes on $AA_{\pm}$ orbitals. In contrast, the wave functions of the $n^1$ and $n^3$ groups have unequivalent contributions on all eight orbitals. Furthermore, they are distorted as a result of the LL anti-crossing. We observe that the wave function nodes are located in specific areas in the real-space unit cell. Because the probability of finding a quantized LL is finite at certain localization centers within a unit cell where the atomic orbitals are located. For monolayer graphene under a B field, the reduced B-field-extended unit cell is a rectangle consisting of 4 C atoms at 1/6, 2/6, 3/6, and 4/6 of the unit cell along the x axis (extension direction of the unit cell in the Landau gauge) \cite{PRBShih}. This is also applied to bilayer graphene. For magic-angle TBLG, the lattice includes bilayer graphene regions with AA and AB/BA stackings and kagome structure. These regions correspond to the dominated orbitals contributing to each band as well as LLs (see Table 1 for details). As a result, the $n^2$ group is dominated by $AA_{\pm}$ whereas the $n^1$ and $n^3$ groups are governed by the combination of all orbitals (see Fig. 1b). This leads to the difference in the localization of LL wavefunctions between them. In particular, there exist localization centers at 1/6, 2/6, 3/6, and 4/6 of the total length of the unit cell for the $n^2$ group, and 1/2 for the $n^1$ and $n^3$ groups. The difference in localization centers forbids the inter-group inter-LL transitions between the $n^2$ group and others. This feature is closely related to the behavior of the plasmon modes, which we discuss below.

\medskip
\par
\begin{figure}[htbp]
\begin{center}
\includegraphics[width=0.9\linewidth]{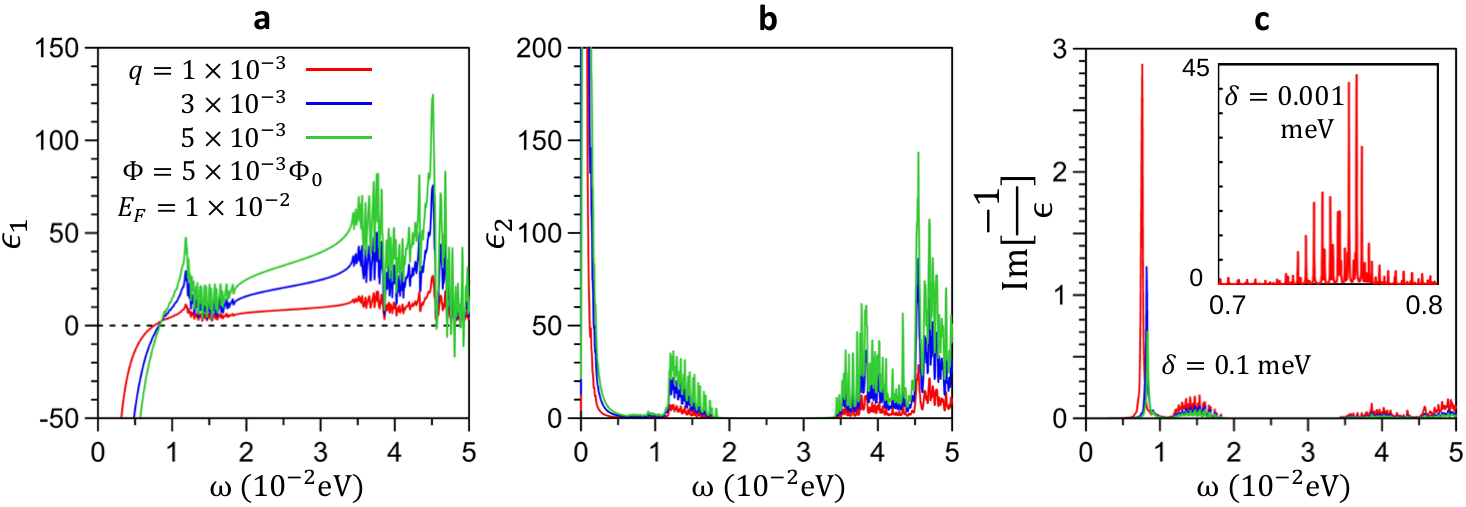}
\end{center}
\caption{(color online) {\bf a} Real part, {\bf b} imaginary part, and {\bf c} the energy loss function of TBLG at magic angle $\theta = 1.1^{\circ}$ under a B field $\Phi = 0.005 \Phi_0$. We present the calculated results for the $n^1$ LL group by choosing $E_F$ = 0.01 eV and various $q$'s. The inset of {\bf c} shows the zoom-in of the significant peaks of the loss function for smaller broadening factor $\delta = 0.001$ meV.}
\label{Fig3}
\end{figure}

The dielectric function, which describes the dynamical screening properties of a system in response to an external electric field, plays an important role in the study of electronic excitations. The dielectric function consists of a real part $\epsilon_1$ and an imaginary part $\epsilon_2$.  These two components are related via the Kramers-Kronig relations. The single-particle excitations (SPEs) can be determined only by $\epsilon_2$ whereas the collective excitations are determined by both $\epsilon_1$ and $\epsilon_2$. Note that $\epsilon_1$ can be vanishing at certain frequencies and this is a necessary condition for the existence of the plasmon modes. At a zero point of $\epsilon_1$, the LL damping of a possible plasmon mode by SPEs can be either weak or strong, corresponding to low or high peak intensity of $\epsilon_2$, respectively. On the other hand, the energy loss function, defined as Im[-1/$\epsilon$], describes the energy transferred from an impinging beam of electrons by the frictional force acting back on it via the induced potential generated by the external electric field \cite{TSO}.  Theoretical results of the energy loss by using this formula are in good agreement with  measurements by  electron energy loss spectroscopy (EELS) for graphene and other materials \cite{R22, R23, R24, R25, R26}. The peaks in Im[-1/$\epsilon$] define the both SPEs and collective modes arising from the electron-charge particle interaction. In order to clearly understand the magnetoplasmons associated with the two flat bands near E = 0, we present the relevant $\epsilon_1$, $\epsilon_2$, and the energy loss function Im[-1/$\epsilon$] in Figs.\,\ref{Fig3}a through \,\ref{Fig3}c. We observe multiple excitation peaks like oscillations in the $\epsilon_1$ and $\epsilon_2$ plots, ranging from $\omega \sim$ 10-20 meV and 35-50 meV. Similar behavior also occurs in the loss function in those energy ranges. Therefore, it is clear that those peaks are attributed to the SPEs between the $n^2$ LLs (for chosen $E_F$ = 0.01 eV). Such oscillation-like peaks correspond to a large number of LLs within a narrow energy range which are quantized from the two flat bands with very high DOS. At a critical $\omega \sim$ 10 meV where the zero point of $\epsilon_1$ appears, $\epsilon_2$ has no contribution. This implies sufficient condition for the collective excitation to take place. There, Im[-1/$\epsilon$] presents a significant peak. Such a peak becomes a group of sharper peaks when a smaller dephasing factor $\delta$ is taken into consideration, referring to the insert of Fig. 3c. This demonstrates the existence of multiple plasmon modes within a narrow range of frequency. The critical frequency $\omega$ where $\epsilon_1$ vanishes remains unchanged for various $q$'s with $q \ge$ 1 $\AA^{-1}$ (see Fig.\,\ref{Fig3}a), indicating the $q$-independence of inter-LL transition. This is because the dispersion of LLs does not depend on the momentum space. On the contrary, the amplitude of the Im[-1/$\epsilon$] high peak presents an inverse relationship with $q$ (see the zoom-in of Fig.\   \,\ref{Fig3}c).

\medskip
\par
\begin{figure}[htbp]
\begin{center}
\includegraphics[width=0.9\linewidth]{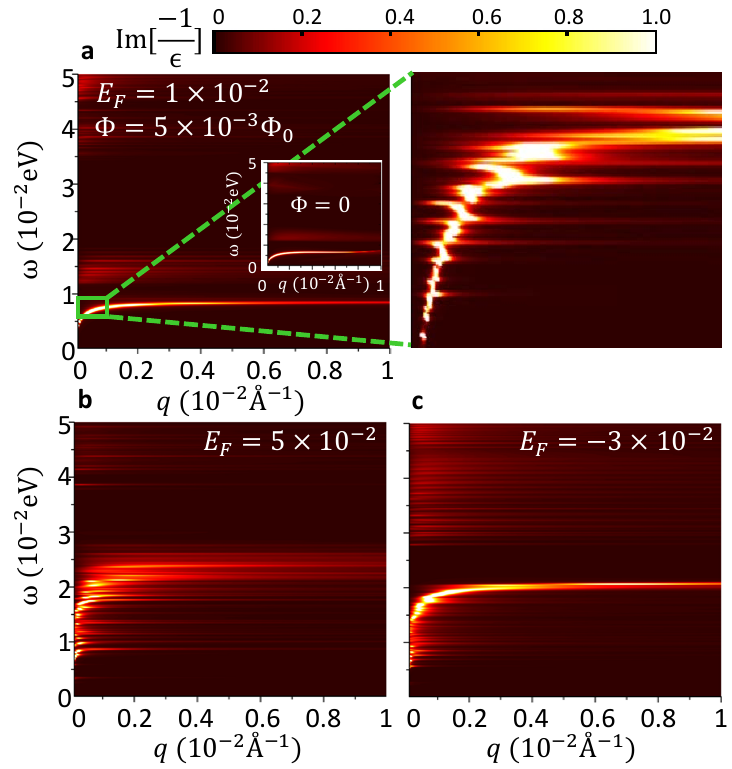}
\end{center}
\caption{(color online) The (q,$\omega$)-dependent phase diagrams for various Fermi energies of TBLG for a magic angle $\theta = 1.1^{\circ}$ under a magnetic flux $\Phi = 0.005 \Phi_0$. The insert in (a) displays the phase diagram at zero magnetic field. The scale bar shows the spectral weight from low (dark) to high (bright).}
\label{Fig4}
\end{figure}

The electronic excitation spectrum is particularly useful for investigating plasmon modes. It is defined as the non-zero spectral weight area in the energy-momentum plane. The electron-electron interactions give rise to coherence of the elementary excitations, leading to a transfer of spectral weight from the SPEs to plasmon modes. An applied B field can modify the plasmons by reorganizing electrons into cyclotron motion. Here we discuss extensively the $q,\omega$-dependent phase diagram to gain insight into the magnetoplasmons of MATBLG. For monolayer graphene, the influence of a B field leads to multiple branches of magnetoplasmons corresponding to a depolarization shift from the cyclotron frequency. Furthermore, there exist the pseudo-quantization of the momentum along a single magneto-exciton due to the nodes of the electronic wave function \cite{layered,GG1,GG2}. For MATBLG, the spectrum associated with the flat bands ($n^2$ LL group) consists of only a single-pronounced magnetoplasmon along a wide range of $q$ (see Fig.\,\ref{Fig4}a), different form the multiple discontinuous plasmon modes of monolayer graphene. Such single-like magnetoplasmon is actually the merging of multiple individual magnetoplasmons with similar frequencies (the zoom-in plot of Fig.\,\ref{Fig4}a), corresponding to the intra-group inter-LL transition within a small energy range of the $n^2$ group (blue lines in Fig. 2a). This is consistent with the feature in the loss function as discussed above. The fact that no magnetoplasmon is found in the nearby spectrum region can be explained by the forbidden inter-group inter-LL transition associated with the $n^2$ group. Besides, the absence of pseudo-quantization phenomenon in TBLG is attributed to the extremely small separation of electronic excitations within high DOS area. In addition, the weight of the magnetoplasmon modes is very strong at low $q$ and faded away for increasing $q$, where they are damped out by the SPEs. At small $q$ ($\le$ 1 $\AA^{-1}$), the critical $\omega$ and $q$ of plasmon establish a proportional relationship. The electron excitation yields horizontal mode with weaker spectral weight at larger $q$, which is the characteristic of the SPEs of dispersionless LLs.
Interestingly, the single-like magnetoplasmon resembles the nearly-flat plasmon mode at zero magnetic field, referring to the insert of Fig.\,\ref{Fig4}a. This is because both the quantized LLs and zero-field energy bands are flat and confined in a narrow energy range. The zero-field horizontal plasmon modes were also reported in previous studies for MATBLG \cite{R14, R15}.

\medskip
\par
\begin{figure}[htbp]
\begin{center}
\includegraphics[width=0.9\linewidth]{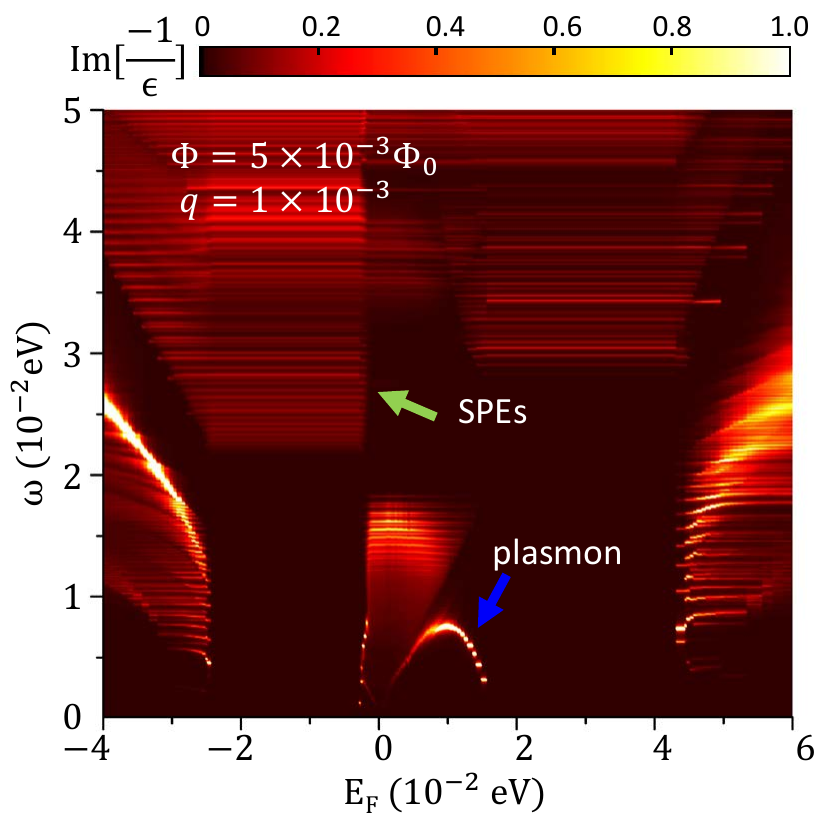}
\end{center}
\caption{(color online) $E_F$-dependent electron excitation spectrum of TBLG at a magic angle $\theta = 1.1^{\circ}$ under a magnetic flux $\Phi = 0.005 \Phi_0$. The scale bar shows the spectral weight from low (dark) to high (bright). }
\label{Fig5}
\end{figure}

The electronic excitations of the $n^1$ and $n^3$ groups are revealed at chosen $E_F$'s, as illustrated in Figs.\,\ref{Fig4}b and \,\ref{Fig4}c. For the $n^1$ spectrum ($E_F$ = 0.05 eV), there are many single continuous excitation modes with strong spectral weight at small $q$ and weaker weight for increasing $q$. These modes are comprised of both magnetoplasmons and SPEs, accompanied by Landau damping and ultimately leading to finite lifetimes of the modes. In general, there is no undamped modes. Similar feature occurs for the $n^3$ spectrum ($E_F$ = -0.03 eV), as shown in Fig.\,\ref{Fig4}c.  However there exists a significant magnetoplasmon at frequency $\omega \approx 20.0$ meV. Such magnetoplasmon is due to the merging of multiple nearby modes associated with the inter-LL transitions occurring in the high density of Landau states (green lines in Fig. 2a). Both the $n^1$ and $n^3$ magnetoplasmons are interspersed by many horizonal SPE modes, different from the single-mode-like excitation of the $n^2$ group.

\medskip
\par
To demonstrate the significant doping-dependence of the electron excitations, we present the ($E_F$, $\omega$)-phase diagram, as shown in Fig.\,\ref{Fig5}. Both SPE and magnetoplasmon modes are clearly displayed in the spectrum. They can be identified based on the plots for $\epsilon_1$, $\epsilon_2$, and the loss function shown in Figs.\,\ref{Fig3}a through \ref{Fig3}c. The intra-group and inter-group SPEs take place at lower and higher frequencies, respectively. The SPEs of dispersionless LLs appear on the horizontal red segments in the spectrum (see the green arrow). We clearly  observe pronounced magnetoplasmons for specific values of $E_F$ at low frequency $\omega$, which is associated with the intra-group LL excitations. These plasmon modes are strongly Landau damped by the SPEs in certain frequency regions where the plasmons undergo Landau damping. The collective excitations of the flat bands yield special magnetoplasmon modes which are well separated from the nearby SPEs (see the blue arrow). This is consistent with the prediction of the strong electron-electron interaction nature of the flat bands which leads to the dominant role of the collective excitations. For the electron excitation of the $n^1$ and $n^3$ LL groups, the SPEs and magnetoplasmon modes are interspersed when $E_F$ enters the LL anti-crossing regions. This implies the occurrence of a complex Landau damping phenomenon. The magnetoplasmon modes are expected to be strengthened for lower or higher $E_F$'s where the collective excitation of LLs with well-behaved wave functions are allowed, e.g., the pronounced plasmon of $n^3$ LL group for $E_F \leq -30$ meV.

\medskip
\par
\begin{figure}[htbp]
\begin{center}
\includegraphics[width=0.9\linewidth]{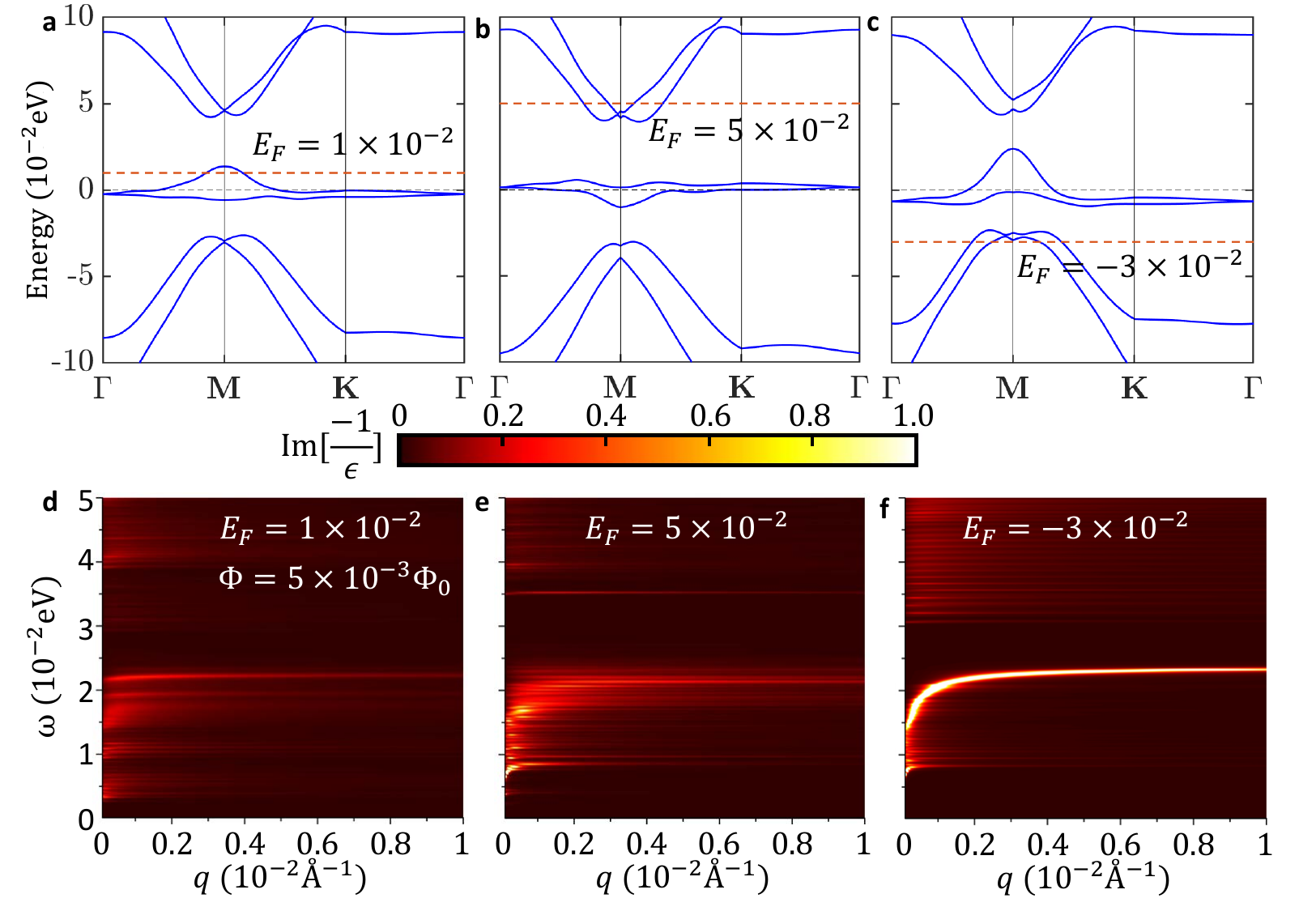}
\end{center}
\caption{(color online) The (q,$\omega$)-dependent phase diagrams for various Fermi energies and corresponding doping-induced potentials of TBLG for a magic angle $\theta = 1.1^{\circ}$ under a magnetic flux $\Phi = 0.005 \Phi_0$. The scale bar shows the spectral weight from low (dark) to high (bright).}
\label{Fig4}
\end{figure}

Now we turn our attention to the effects of modulation of the charge density on the band structures and magnetoplasmon modes. Changing the doping level gives rise to considerable modification of the energy bands \cite{dopingband1, dopingband2, dopingband3} and other electronic properties of MATBLG \cite{dopingelec1, dopingelec2}. The apparent qualitative changes in the band dispersion of doped MATBLG regarding both the two flat bands and other bands away from the neutrality point are shown in Figs. 6(a) through 6(c). For the two nearly flat bands, we observed the variation of band width, the substantial shift of bands, and flattening of the conduction band upon electron doping. These significant alterations of the band structures due to doping are in good agreement with previous reports \cite{dopingband1, dopingband2}. The changes in energy dispersion and $E_F$ caused by doping are expected to enrich the magnetoplasmons of MATBLG.

\medskip
\par
Electronic excitations of MATBLG can be modified through the doping-induced screened potential and change in $E_F$. The $(q,\omega)$-dependent phase diagram changes significantly at low doping level, as shown in Fig. 6(d) for $E_F$ = 0.01 eV. The spectral structures are associated with the intragroup excitations within the $n^2$ LLs quantized from the two flat bands. By considering the nonvanishing doping-induced potential, the single-pronounced magnetoplasmon with strong spectral weight no longer exists. Our calculation shows that such collective excitations are damped out by SPEs. As a result, the $(q,\omega)$-dependent phase diagram displays multiple SPEs (see Fig. 6(d)). On the other hand, for the higher electron and hole doping levels, the induced potential do not lead to a noticeable variation on the spectral features. Only a slight change in spectral weight of the magnetoplasmon modes is observed, referring to Figs. 6(e) and 6(f).

\section{Summary and Concluding Remarks}
\label{sec4}

We have presented a detailed theoretical investigation concerning the electronic excitations for MATBLG in the presence of an external ambient magnetic field. We constructed the magnetic Hamiltonian by using the minimal-basis TBM in conjunction with the Peierls substitution. The eigenvalues and wave functions were inserted into the dielectric function to calculate the electron excitation spectra. We found pronounced magnetoplasmons associated with the intra-group inter-LL excitations. These plasmon modes can be significantly modified by doping and transferred momentum. We found exotic electron excitation spectrum associated with the unique quantized LLs, including pronounced magnetoplasmon and flat SPE modes as well as Landau damping phenomenon. Especially, the collective excitation associated with the flat bands leads to a single significant magnetoplasmon mode. Such exotic magnetoplasmon has never been observed in graphene and other 2D systems. From many points of view, TBLG is a unique system and its applications could involve microelectronics technologies, integrated circuits and devices, chemical sensors, ultrafast modulators and high mobility transistors, as well as the use of excitons as qubits for quantum computing. Our predicted results can lead to possible application of MATBLG in the field of magneto-plasmonics.

\section*{Acknowledgement(s)}
The authors would like to thank the MOST of Taiwan for the support through Grant No. MOST111-2811-M-006-009.
G.G. would like to acknowledge the support from the Air Force Research Laboratory (AFRL) through Grant No. FA9453-21-1-0046.

\end{document}